Getting the Swing of Surface Gravity

Brian C. Thomas & Matthew Quick

Department of Physics and Astronomy
Washburn University
1700 SW College Ave.
Topeka, KS 66621
brian.thomas@washburn.edu

Sports are a popular and effective way to illustrate physics principles. Baseball in particular presents a number of opportunities to motivate student interest and teach concepts. Several articles have appeared in this journal on this topic[1], illustrating a wide variety of areas of physics. In addition, several websites[2] and an entire book[3] are available. In this paper we describe a student-designed project that illustrates the relative surface gravity on the Earth, Sun and other solar-system bodies using baseball. We describe the project and its results here as an example of a simple, fun, and student-driven use of baseball to illustrate an important physics principle.

This project was completed to satisfy a course requirement in an introductory astronomy course at Washburn University (a Masters-level university) in Topeka, Kansas. The assignment was an open-ended, independent project designed and executed by the student. The requirements were that the project must be self-designed and related to astronomy, with creativity emphasized.

The project described here asks the question "What would it be like to play baseball on other planets?" Two quantities were chosen for comparison: "hang time" of the baseball and distance from home plate to the center field fence. These values are affected by the *surface gravity* of the planet or other body. Surface gravity means the gravitational acceleration at the surface of the body, which depends on both the body's mass and the distance from the center to the surface. We realize that one would not actually be able to stand (let alone play baseball!) on the surface of a planet such as Jupiter; the idea is to help students understand surface gravity. This concept may be difficult for some students, since it involves variation of two parameters simultaneously. Therefore, we hope this exercise will be both engaging and useful in helping students understand the counter-intuitive fact that even a planet with *greater* mass than the Earth (for instance, Neptune) may have a *smaller* surface gravity, or vice versa.

For the student project presented here, empirical data was used to determine the hang time (that is, how long the ball was in the air after being hit). The student and two friends recorded the flight time of 50 hits each. Each batter's times were averaged and then these three were averaged to get a final hang time value of 1.2 seconds. This value is an average over *all* hits, regardless of distance. A more typical value for a fly ball to the outfield is around 3 or 4 seconds. One can either use our approach (measure and average over all values) or only use values for hits that reach beyond a certain distance. The computed results will of course differ based on what data is used.

To find the hang time on other planets, the Moon, and the Sun, the measured value on Earth was divided by the *relative* gravitational acceleration at the surface of each solar system body. Relative gravitational acceleration is simply the ratio of the acceleration due to gravity on the surface of the body (e.g. the Sun) to that on the Earth's surface. A larger gravitational acceleration leads to shorter hang times since the ball is accelerated back towards the surface more strongly. The advantages of using this relative quantity are that it is simple to use; it does not require any detailed first-principles calculations; and it helps students understand how the values compare between the objects being considered. We feel this comparative value is more helpful to students than absolute numbers would be, and the point of the exercise is to illuminate the relative differences between surface gravities. An instructor can, of course, extend the exercise to include more explicit first-principles calculations if desired.

Values for relative gravitational acceleration were obtained from a webpage hosted by the University of Virginia Physics and Astronomy Department.[4] Other sources exist, of course, including tables in most introductory astronomy textbooks. An instructor could also have students calculate the acceleration due to gravity on each solar system body using Newton's gravitational acceleration equation,

$$g = \frac{Gm}{r^2},$$

where $g$ is the acceleration due to gravity, $G$ is Newton's universal gravitational constant (6.67 x $10^{-11}$ N m$^2$ kg$^{-2}$), $m$ is the mass of the body, and $r$ is the radius of the body.

The method described above was also used to calculate the distance from home plate to the center field fence. The distance at Kauffman Stadium in Kansas City, Missouri (our nearest professional baseball field) was used as the "standard" distance, which was divided by the relative gravitational acceleration at the surface of each body. Table I gives values for hang time and ball park size (computed using our measured hang time and the size of Kauffman Stadium, respectively) for the Sun, the Moon, the classical planets, and one dwarf planet. One could of course use any ball park size or hang time value for this calculation, which would change the table values accordingly.

There are some potential misunderstandings that students may have about the topics in this activity. For instance, it is common to confuse the force of gravity with the acceleration due to gravity; instructors should point out (preferably with examples) that the acceleration does not depend on the mass of the baseball, but instead depends on the size and mass of the planet or other solar system body. An instructor could also have the students measure the mass of the ball on Earth and compute its weight here and on the other solar system bodies. Also, students may have difficulty understanding the concept of relative gravitational acceleration; the instructor should have the students use units carefully, noting that this relative value is a ratio of accelerations and is therefore dimensionless. This could also be an opportunity to point out the utility of dimensionless quantities in physics.

The comparisons presented here can help students understand the relative surface gravity on the various major bodies in the Solar system. Particularly surprising for many students may be the result that the hang time and field size on Saturn, Uranus and Neptune are quite similar to that on the Earth. This can be used to illustrate the fact that surface gravity depends not only on the mass of the body but also its size (radius). Similarly, it may be instructive to have students investigate how the acceleration changes if the ball field is, for instance, on top of Mount Everest, or at the bottom of the Valles Marineris on Mars.

This exercise, like any involving topics that students are interested in, has the potential to hold students' attention and engage them in understanding what may be a confusing topic. We hope that other physics and astronomy instructors will find this project, or at least its results, useful in their own classrooms. Instructors may want to expand or extend this activity, maybe including other sports that their students have particular interest in. One could also have students design their own project based on this idea; there is plenty of room for creativity and open-ended inquiry, which we know both engages students and helps them to better understand the concepts we seek to teach.

Acknowledgements
The first author thanks Todd McAlpine (University of Findlay) for the original idea of an independent student research project in introductory astronomy. The second author thanks Kaylan Thompson, Adam Riley and Bentley Johnson for assistance with data collection.

Table I – Hang times and ball park sizes for student-produced values presented here.

| Solar System Body | Hang time (seconds) | Ball park size (feet) |
|---|---|---|
| Sun | 0.04 | 14 |
| Mercury | 3.2 | 1084 |
| Venus | 1.3 | 452 |
| Earth | 1.2 (measured) | 410 (Kauffman Stadium) |
| Moon | 7.2 | 2469 |
| Mars | 3.2 | 1087 |
| Jupiter | 0.50 | 173 |
| Saturn | 1.3 | 447 |
| Uranus | 1.3 | 461 |
| Neptune | 1.1 | 366 |
| Pluto | 20.2 | 6949 |